\documentclass[12pt,preprint]{aastex}

\begin{document}

\title{HST and Optical Data on SDSSJ0804+5103 (EZ Lyn) One Year 
after Outburst\footnote{Based on observations 
made with the NASA/ESA Hubble Space
Telescope, obtained at the Space Telescope Science Institute, which is
operated by the Association of Universities for Research in Astronomy, Inc.,
(AURA)
under NASA contract NAS 5-26555, and with the Apache Point Observatory 3.5m 
telescope which is owned and operated by the Astrophysical Research
Consortium}}

\author{Paula Szkody\altaffilmark{1},
Anjum S. Mukadam\altaffilmark{1},
Edward M. Sion\altaffilmark{2},
Boris T. G\"ansicke\altaffilmark{3},
Arne Henden\altaffilmark{4},
Dean Townsley\altaffilmark{5}}
\altaffiltext{1}{Department of Astronomy, University of Washington, Seattle, WA 
98195; szkody@astro.washington.edu, mukadam@astro.washington.edu}
\altaffiltext{2}{Department of Astronomy \& Astrophysics, Villanova University,
Villanova, PA 19085; edward.sion@villanova.edu}
\altaffiltext{3}{Department of Physics, University of Warwick, Coventry
CV4 7AL, UK; boris.gaensicke@warwick.ac.uk }
\altaffiltext{4}{AAVSO, 49 Bay State Road, Cambridge, MA 02138; arne@aavso.org}
\altaffiltext{5}{Department of Physics \& Astronomy, University of Alabama,
Tuscaloosa, AL 35487; Dean.M.Townsley@ua.edu}

\begin{abstract}
We present an ultraviolet spectrum and light curve of the short orbital
period cataclysmic variable EZ Lyn obtained with the
Cosmic Origins Spectrograph on the Hubble Space Telescope 14 months after
its dwarf nova outburst, along with ground-based optical photometry.
The UV spectrum can be fit with a 13,100K, log g=8 white dwarf using 0.5 solar
composition,
while fits to the individual lines are consistent with solar abundance for
Si and Al, but only 0.3 solar for C.
The Discrete Fourier Transforms of the UV and optical light curves at 14 months 
following outburst show a prominent period at 256 sec. This is the same period
reported by Pavlenko in optical data obtained 7 months and one 
year after outburst, 
indicating its long term stability over several months, but this period is not 
evident
in the pre-outburst data and is much shorter than the 12.6 min period 
that was seen in observations obtained during an interval from 8 months to 
2.5 years after the 2006 outburst. In some
respects, the long and short periods are similar to the behavior seen in GW Lib
after its outburst but the detailed explanation for the appearance and
disappearance of these periods and their relation to non-radial pulsation
modes remain to be explored with theoretical models.
\end{abstract}


\section{Introduction}

Among the new cataclysmic variables found in the Sloan Digital Sky Survey
was the $g=17.9$ mag system SDSSJ080434.20+510349.2 (Szkody et al. 2006), which
has since received the name EZ Lyn. The SDSS spectrum
showed Balmer emission lines with deep central absorption (indicative
of high inclination) and surrounded by increasing broad absorption of
the higher order Balmer lines (indicative of a large contribution of the
white dwarf to the optical light). Time-resolved spectra showed an orbital
period of 85 min, near the orbital period minimum (G\"ansicke et al. 2009), 
while a 5 hour light curve showed a periodic modulation at
half this period, which began with a peak-to-peak amplitude of 0.05 mag
and then increased to 0.2 mag when the object suddenly underwent a
brightening from 18.2 to 17.7 in blue light. No short period variations were
evident in the Discrete Fourier Transform (DFT) at that time (2005 Jan). 
Analysis of
further photometry (Zharikov et al. 2008, Kato et al. 2009) revealed a shallow 
eclipse of 0.05 amplitude. In 2006 March, Pavlenko et al. (2007) found the 
object undergoing a dwarf nova outburst at 13th mag, 
and another outburst at 12.5 mag was observed
 in 2010 September (Kato et al. 2012).
Long term photometry following the 2006 outburst and throughout the 
quiescent interval revealed a periodicity at 12.6 min
that first appeared 8 months after the 2006 outburst  
and lasted for about 900 days (Pavlenko 2009, Zharikov et al. 2013).
This period was interpreted as possible
non-radial pulsations of the white dwarf. After
the 2010 outburst, this period was not evident for 11 months. However, a
shorter period at 4.28 min (257s) was reported by Pavlenko et al. (2012)
during 2011 April and September (7 months and 1 yr) after this outburst.

The instability strip for the accreting white dwarfs that are pulsating
has been found to be wider (10,500-16,000K;
 Szkody et al. 2010) than the narrow strip
(10,800-12,300K; Gianninas et al. 2005) for non-interacting DAV pulsators.
Arras et al. (2006) suggest this is due to the metal-enriched atmospheres
of the accreting white dwarfs, which enable He ionization zones as well
as H. Their models show that if the He abundance is $>$ 0.38, there is a
hot instability strip near 15,000K due to \ion{He}{2} ionization. The cool
H ionization strip merges with this one for a He abundance $>$0.48.
Since it is well known that dwarf novae are heated by their outbursts and
subsequently cool on timescales of several years
(G\"ansicke \& Beuermann 1996, Sion, 1999; Piro et al. 2005; 
Godon et al. 2006), the dwarf novae that
contain pulsating white dwarfs can provide
a unique opportunity to observe a white dwarf that moves out of the strip
during the outburst heating and then moves back in as it cools, on timescales
much shorter than evolutionary ones for non-interacting pulsators. Systems
such as GW Lib and V455 And show this expected type of behavior, with their
quiescent pulsations disappearing for a few years after outburst and then
starting to recommence at different periods (van Zyl et al.
2004, Bullock et al. 2011, Szkody et al. 2012, Silvestri et al. 2012).
However, the behavior of EZ Lyn is opposite, showing possible pulsations
only in the 1-2 years after outburst and not during quiescence. It raises
the possibility that this system normally sits outside the strip at quiescence, 
then moves through
it after its outburst heating.

While the optical
spectra and the orbital period of EZ Lyn are very similar to those of the 
other accreting pulsators, EZ Lyn is unique with its low superoutburst
amplitudes and its short interval between outbursts. 
The other known
systems are similar to GW Lib, which has an outburst amplitude near 9 mag 
and an outburst recurrence time of
20-30 yrs, behavior that is in accordance with
disk instability models for ultrashort period dwarf novae (Howell et
al. 1995). In addition, EZ Lyn is also peculiar in having sporadic
sharp increases in brightness of $\sim$0.5 mag during quiescence, sometimes
termed mini-outbursts (Szkody et al.
2006, Pavlenko 2009; Pavlenko et al. 2012). This type of sudden increase in
brightness has also been observed in the system SDSSJ1238-0339 (Szkody et al.
2003, Aviles et al. 2010) which does not contain a pulsating white dwarf.

Since the temperature of the white dwarf is best determined in the
ultraviolet, using the Ly$\alpha$ region (G\"ansicke et al. 2005), and the pulsations are
usually strongest in the ultraviolet as well for low $\ell$ modes
(Robinson et al. 1995, Nitta et al. 2000, Szkody et al. 2007),
we pursued Hubble Space Telescope (HST) spectra to determine the
properties of the white dwarf in EZ Lyn. The 2010 outburst happened
after the proposal was accepted, so the observation was obtained
as late as possible in the HST cycle in order to maximize the time for the 
white dwarf
to cool after its outburst. Ground-based observations were conducted
1.5 months after outburst and also coordinated with the HST data that
were obtained 14 months
after outburst.

\section{Observations}

The UV data were obtained during two consecutive HST
 orbits with the Cosmic Origins Spectrograph
(COS) and the G140L grating setting of 1105 on 2011 November 3. This grating 
enables coverage
of 1130-2000\AA\ with a resolution of $\sim$0.75\AA. The spectra were
obtained in time-tag mode so that light curves could be easily constructed. The
analysis used PyRAF routines from STSDAS package hstcos (version 3.14).
Various extraction widths were tried to optimize the S/N, with the best
value found for a width of 41 pixels. This extraction was then applied to the
summed spectrum from the two orbits.

An UV light curve was created from the time-tag data by summing the fluxes over
the good S/N area of 1136-1785\AA\ (after deletion of the strong geocoronal
emission lines at 1200 and 1300\AA) using 3 sec bins. This light curve
was then converted to fractional amplitude by dividing by the mean and
subtracting one for ease in computing a Discrete Fourier Transform (DFT).

Fast time-series optical photometric data were obtained on 2010 November 8
 and a 
year later on 2011 November 5,7 during 2 nights that
lagged the HST observations by 1.5 and 3.5 days (Table 1). The observations
were accomplished using the 3.5m telescope at Apache Point Observatory (APO),
with the  frame transfer CCD Agile (Mukadam et al. 2011) and a BG40
filter.
Measurements 
surrounding the
outburst and the
HST dates were also provided by the AAVSO network and are available from
their archive{\footnote{http://www.aavso.org/data-access}}. 
The AAVSO observations show that EZ Lyn had returned to a visual magnitude
close to
quiescence (V$\sim$17.8) by 2010 October 28 ($<$ 1.5 months after outburst).
This value continued throughout the time of the HST observations.

\section{UV Spectrum}

The averaged and smoothed (by 3 point boxcar) COS spectrum obtained
on 2011 November 3, one year after outburst, is shown in Figure 1.
Except for weaker CIV1550 emission and deeper metal absorption lines, the 
spectrum is very similar to the dwarf 
nova HV Vir (Szkody et al. 2002).
The broad absorption features near 1400 and 1600\AA\ are the quasi-molecular
Ly$\alpha$ absorption of H$_{2}^{+}$ and H$_{2}$ that are visible at
temperatures below $\sim$18,000K and $\sim$13000K respectively. Besides
the SiII lines at 1522,1533\AA\ and the AlII line at 1670\AA, the other
strong absorption lines are 
CI (1355,1431,1463,1467,1561,1657\AA) and CII (1323,1335\AA). There is
weak emission at 
CIII (1175\AA), CIV (1550\AA) and HeII(1640\AA). Figure 2 shows an expanded 
figure with
the best fit white dwarf (temperature of 13,100K, log g=8, 0.5 solar
abundance )
obtained from a grid of models (Hubeny \& Lanz 1995). This temperature places 
it in the middle of the instability strip for accreting pulsating white dwarfs
if this temperature is the quiescent one. While some short period dwarf
novae are known to take more than 3 years to cool from outburst to 
quiescence (Godon et al. 2006; Szkody et al. 2012), those systems have
very large outburst amplitudes (8-9 mag) and long recurrence times
(20-30 years). The outbursts of EZ Lyn are smaller amplitude (5 mag)
and occur more frequently (4 yrs). Thus, the heating should be less and the
corresponding cooling time shorter. This is corroborated by the fact that
GW Lib remained 0.5 mag above its quiescent value for several years post
outburst while EZ Lyn returned to optical quiescence in a few months.
Thus, we expect the temperature determined from the COS data to be very
close to the quiescent value. 

The prominent presence of  CI lines in the FUV spectrum of EZ Lyn resembles
the line spectra of the white dwarfs in WZ Sge and HV Vir during quiescence,
which have roughly the same surface temperature as EZ Lyn. The line feature at
1355\AA, also present in HV Vir (Szkody et al. 2002), is presumed to be due to
a transition of CI. In HV Vir, the increased carbon abundance required to
successfully fit the strong CI 1355\AA\ feature
curiously led to line strengths for all of the other CI lines far in excess
of what is observed. However, in HV Vir, the
observed CI 1355 feature is considerably stronger than seen in EZ Lyn. In
order to have a closer look at the chemical abundances for the individual
elements C, Si and Al in the accreted atmosphere of EZ Lyn, we used the 
T$_{eff}$,
and log g previously determined to generate model fits for a range of chemical
abundances 0.01 , 0.1 , 0.2 , 0.3 , 0.5 , 0.7 and 1.0 times solar. The best
overall fits to the individual lines reveal that the Si and Al abundances are
$\sim$solar,
while the CI abundance is $\sim$0.3 times solar. Thus, the C
abundance appears to be depressed relative to Si and Al. While a lower C
abundance could be taken as evidence of CNO processing, our spectral range
contains no N feature to determine if the abundance of N is elevated relative to
C. The best overall model fit to the line spectrum occurs for a solar
composition of the accreted atmosphere. This fit is shown in Figure 3.

We also tried to constrain the rotational velocity by a detailed 
fitting of the SiII 1526, 1533 doublet absorption
lines. We tried rotational velocities of 75, 100, 150, 200, 250, 300, 350, 400
and 500 km/s with Si abundances of 0.3, 0.5, 0.7, 0.8, 0.9, 1.0, 1.5, 2.0, and
3.0 times solar, respectively. The best fitting rotational velocity is vsini =
225$\pm$75 km s$^{-1}$ with Z = 1.0 solar. The solar metal abundance that we 
used has an uncertainty of a factor of 2. Thus, for our range of possible Vsini
values between 150km/s and 300km/sec, the corresponding metal abundance Z has a
possible range between Z = 0.5 and Z = 2.0 solar where the lower limit of vsini
is obtained with the
lower limit of 0.5 and the upper limit of vsini is obtained with the upper limit
of Z = 2.0.

\section{UV and Optical Light Curves} 

The light curves created from the COS data, using 3 sec binning, are shown
in Figure 4. While there is obvious variability of $\sim$40\%, there is
no apparent eclipse feature in the UV, but the phase coverage is
incomplete. The eclipse is apparent in each
of the 3 longer optical datasets (Figures 5-7) and the times of the bottom of
the eclipse are given in Table 2. Extrapolating back from the time
of the optical eclipse on 5 Nov (JD 2,445870.9094) with the period of
0.0590048d from Kato et al. (2009) shows that the eclipse would have
occured just prior to the 2nd HST orbit (the 2 orbits cover phases
0.27-0.71 and 0.08-0.71) so the eclipse is not covered. The optical data also 
show the
double-humped variation that is typical of short period systems with
low mass transfer (Patterson et al. 1998, Aviles et al. 2010, Zharikov et al. 2013). The lack of 
this
feature in the UV data is consistent with an origin of most of the UV light
from the white dwarf, which is not eclipsed, and not from the disk, which
is the source of the double-hump modulation.   

To search for periodic features in a quantitative way, DFTs were computed
for all the data and the best fit periods were found by least-squares
fits. To determine the significance of the periods, an estimate of the
3$\sigma$ white noise was determined by shuffling the intensities (with
the periods extracted) to create a pure white-noise light curve. The
average amplitude of the DFT of this light curve then gave a 1$\sigma$ measure
and this was repeated 10 times to provide an empirical 3$\sigma$ limit.
Figures 4-7 show the original DFTs, and the resulting pre-whitened ones
with the periods removed, as well as the 3$\sigma$ limits. Due to the
large low frequency variation present in the data obtained on 2010
November 8 (1.5 months
after outburst), this trend was first removed and the DFTs computed with
the original and the subtracted versions of the light curves.

The DFT of the UV light curves (Figure 4) reveals a high amplitude periodicity
at 256.1 $\pm$0.2 sec (4.27 min) as well as a longer period at 43.2$\pm$0.9 min.
While the longer period could be a result of the short observation length of 
the HST orbits, it is more likely the first harmonic of the orbital period 
(84.97 min) within the error bar, as this period is often present in our
longer optical datasets. The 4.27 min period is the same period reported in 
the Pavlenko et al.(2012) 
optical data obtained on 2011 April 30, September 5 and 6 (7 and 11 months
past outburst). Our optical data obtained 1.5 months after outburst, close
to quiescence (Figure 5), shows variability, but a period at
256 sec is not detected. Our 2011 November optical data obtained 2 and 4 days
after the UV data (Figures 6 and 7) do show 
significant periods consistent with 256 sec (254.1$\pm$1.4s on November 5 
and 255.0$\pm$0.5s on November 7). Thus, this appears to be a stable
feature for over 6 months. The stability rules out a QPO feature. 
If this were the rotation period of the white
dwarf, it would be present during all quiescent data, not merely appear
after outburst. Even more puzzling is why this period is different from
the 12.6 min period that lasted for $\sim$900 days following the 2006 outburst.

The UV/optical ratio of the amplitudes of the 4.27 min period is 66/6=11,
a value typical for non-radial pulsators (Szkody et al. 2007). 
If the 13,000K white dwarf temperature is the quiescent one for EZ Lyn, it
sits squarely in the middle of the instability strip, yet it is not observed
to pulsate at quiescence. A possible explanation could be that it has a relatively
high He abundance so that it is too cool for the \ion{He}{2} ionization strip which exists 
at 15,000K (Arras et al. 2006)
and only moves through the strip after outburst heating.
However, the cessation of the pulsations
at quiescence and the different periods would imply that EZ Lyn is very
close to the stability zone for its mass and composition and small amounts
of heating are enough to displace its stability. This is consistent with
the small amplitude of its outbursts (which would cause less heating).
Pavlenko et al. (2012) also noted that the 12.6 min variation that was
evident during the 900 days after the 2006 outburst varied in amplitude
(0.01-0.03 mag) and period (732-768 sec). This wandering in frequency
appears to be a common trait of accreting pulsators (Uthas et al. 2012,
Szkody et al. 2012)
and it is not clear what causes this effect.

GW Lib is currently the system with the most available data on the pulsating
white dwarf following its outburst (Szkody et al.
2012, Chote \& Sullivan 2013). Even though there are obvious differences in the outbursts and 
in the quiescent pulsation spectrum of the two systems, their behavior after
outburst contains some interesting commonalities. Both exhibit a long
period and a short period that appears to be triggered by mass accretion.
In GW Lib, this period is near 19 min (Copperwheat et al. 2009, Bullock et 
al. 2011, Vican et al. 2011)
vs the 12.6 min in EZ Lyn. In both
of these systems, the long period was observed to appear in the
year following the outburst, when the system had returned to close to its 
quiescent level.
In GW Lib, this period lasted for 4 months, while in EZ Lyn for 2.5 years.
 The short
periods that appear in both UV and optical data when the long period is
not present
are 290 sec in GW Lib (Szkody et al. 2012) and 256 sec in EZ Lyn. This short period was evident
from 3-4.25 years past outburst in GW Lib, and then was replaced by the
return of the 19 min period during the 5th year after outburst (Chote \&
Sullivan 2013).
The short period in EZ Lyn appeared 7 months after the latest outburst
and lasted for at least 6 months. While these periods are generally
incoherent over several nights, the study of the 19 minute period present
in 2012 in GW Lib revealed coherence over 2 nights, with combinations of
frequencies that were consistent with non-radial g-modes (Chote \& Sullivan 2013). 

The thermal timescale at the base of the convection zone dictates
which pulsation modes are excited, so longer periods are expected as the star 
cools and the base of the convection zone sinks deeper in the white dwarf
(Brickhill 1992, Goldreich \& Wu 1999, Wu 2001, Montgomery 2005). If both the 
long and short periods
are related to non-radial pulsation, then the only way to observe short period modes
after the excitation of longer periods, is for the white dwarf to undergo 
accretion-related heating. The change from short to long periods as the
white dwarf cools is consistent with being due to the excitation of a different
mode at lower surface temperatures. However, conclusions about what the mode
frequencies mean about the interior of the star and the development of a 
consistent picture of the changing excitation requires further theoretical work. 

\section{Conclusions}

Our COS spectrum of EZ Lyn obtained in 2011 November, 14 months after its dwarf 
nova outburst,
shows prominent absorption lines from a white
dwarf that is consistent with a temperature of 13,100K, and log g=8. The
lines of Si and Al indicate a metal composition near solar, while the C 
abundance is only 0.3 solar.
The UV and optical light curves reveal a significant periodicicy
at 256 sec, with an expected higher amplitude in the UV than the optical. This
period is identical to that evident in the optical in April and September
2011 (Pavlenko et al. 2012). While this appears to be a non-radial pulsation
that is triggered by the outburst, the difference in behavior from data
obtained after a previous outburst in 2006 (Pavlenko 2009), which showed
a longer period at 12.6 min remains to be explained.

\acknowledgements
We gratefully acknowledge Elizabeth Waagen and 
the observers of the AAVSO for their efforts in
monitoring the brightness state of EZ Lyn that allowed the HST observations
to proceed, especially David Boyd, Michael Cook, William Goff, Robert Koff, 
David Lane,Gordon Myers, and Larry Wade. We also thank Elena Pavlenko for
useful discussions about her data. 
This work was supported by NASA grant
HST-GO12231.01A from the Space Telescope Science Institute, which is
operated by the Association of Universities for Research in Astronomy, Inc.,
for NASA, under contract NAS 5-26555, and by NSF grant AST-1008734.

\clearpage
\begin{deluxetable}{lcllll}
\tablewidth{0pt}
\tablecaption{Summary of Observations}
\tablehead{
\colhead{UT Date} & \colhead{Obs} & \colhead{Instr} & 
\colhead{Filter} & \colhead{UTC Time} & \colhead{Exp (s)}}
\startdata
2010 Nov 8 & APO & Agile & BG40 & 07:11:59-08:36:54  & 5 \\
2011 Nov 3 & HST & COS & G140L & 21:22:58-22:00:34 & Time-tag \\
2011 Nov 3 & HST & COS & G140L & 22:32:27-23:25:06 & Time-tag \\
2011 Nov 5 & APO & Agile & BG40 & 09:39:58-12:02:28 & 30 \\
2011 Nov 7 & APO & Agile & BG40 & 07:59:40-12:16:05  & 30-60 \\
\enddata
\end{deluxetable}

\clearpage
\begin{deluxetable}{lll}
\tablewidth{0pt}
\tablecaption{Eclipse Times}
\tablehead{
\colhead{UT Date} & \colhead{UTC Time}}
\startdata
2010 Nov 8 & 08:30:49 \\
2011 Nov 5 & 09:49:28; 11:14:08 \\
2011 Nov 7 & 09:57:51; 11:22:51\\
\enddata
\end{deluxetable}

\clearpage
\begin{figure}
\figurenum {1}
\plotone{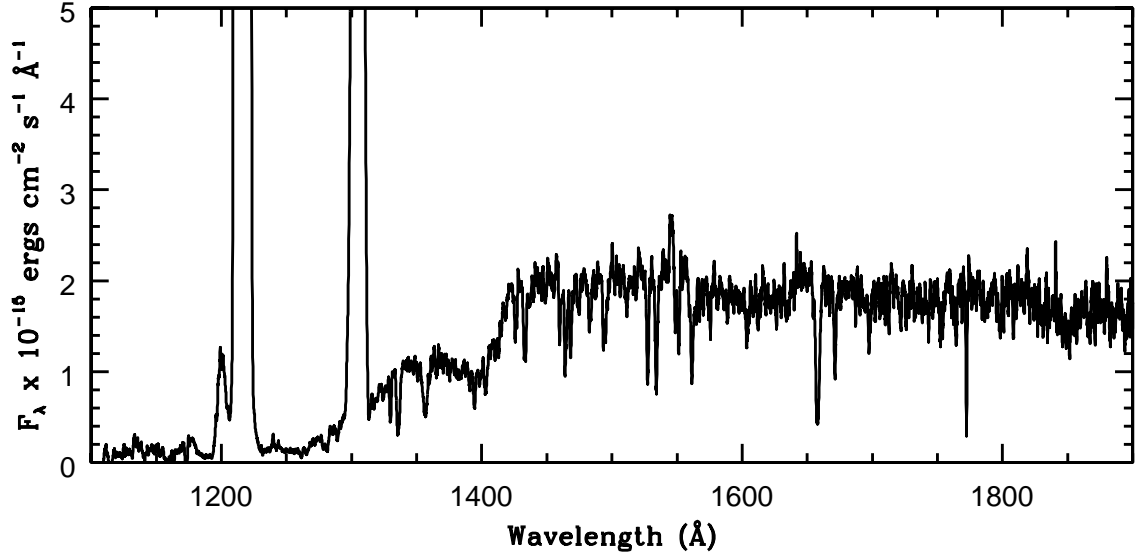}
\caption{COS G140L spectrum 2011 Nov 3 (14 months after outburst) 
smoothed with a 3pt running
boxcar. Strong emission lines are 
airglow from Ly$\alpha$ and OIII.}
\end{figure}
\clearpage

\begin{figure}
\figurenum {2}
\includegraphics[angle=-90,width=7in]{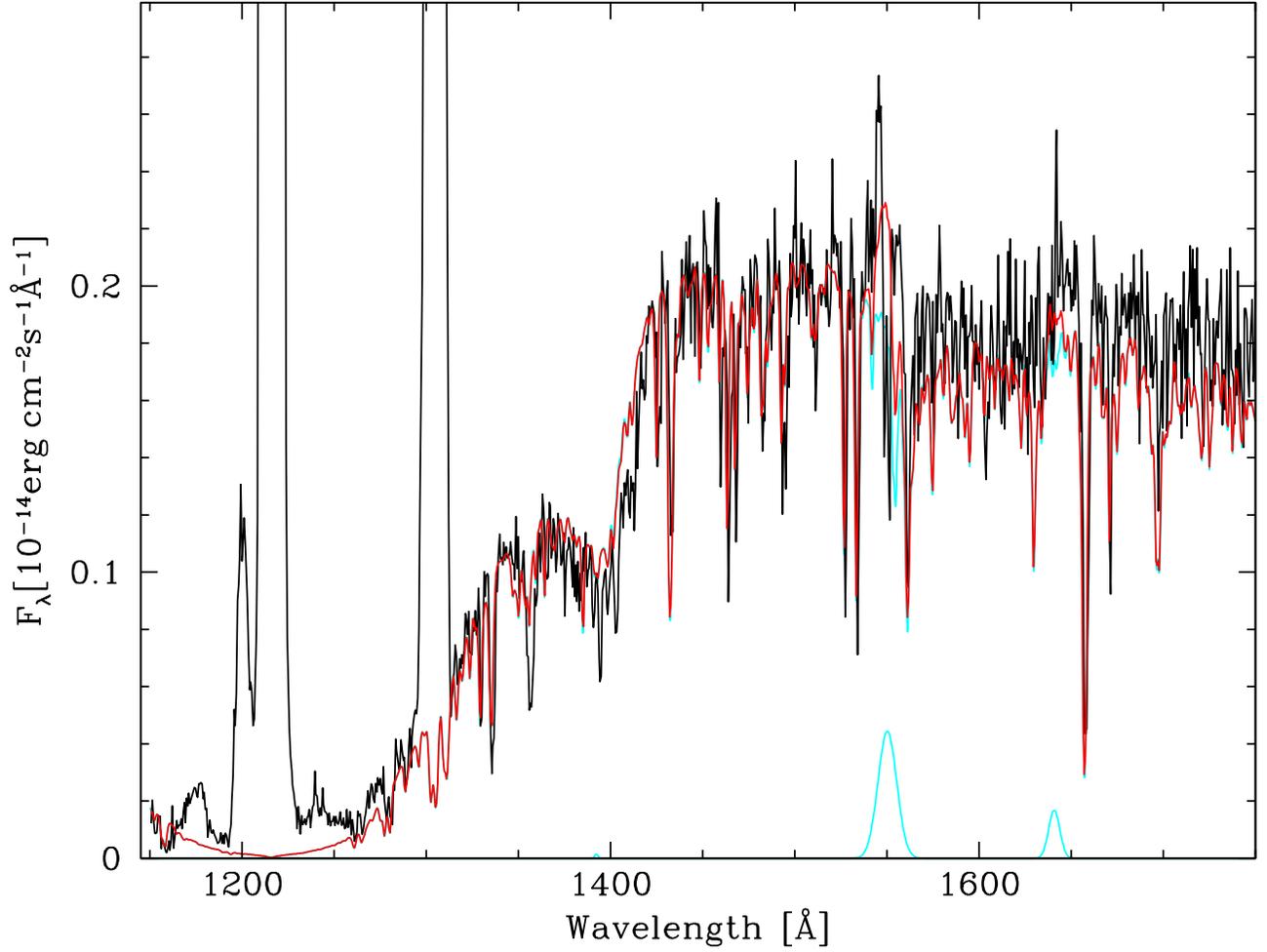}
\caption{COS spectrum (black) fit with a 13,100K, log g=8.0 white dwarf with 0.5 solar
composition (red). CIV and
HeII emission lines are represented by Gaussians (blue).}
\end{figure}

\clearpage
\begin{figure}
\figurenum {3}
\includegraphics[height=8in]{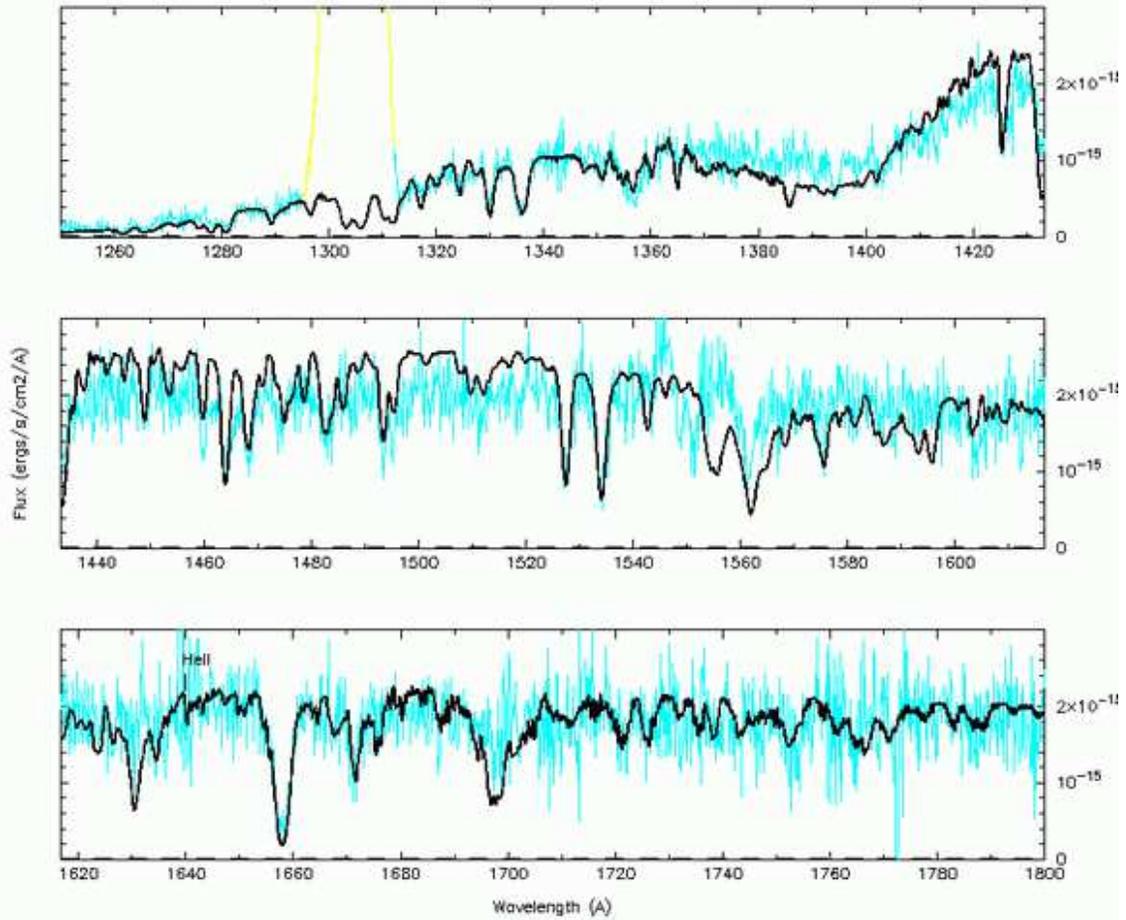}
\caption{COS spectrum fit with a 13,000K,log g=8 solar composition white dwarf.
 Fits to
individual lines yield solar abundcances for Si and Al and 0.3 solar for C.}
\end{figure}

\clearpage
\begin{figure}
\figurenum {4}
\plotone{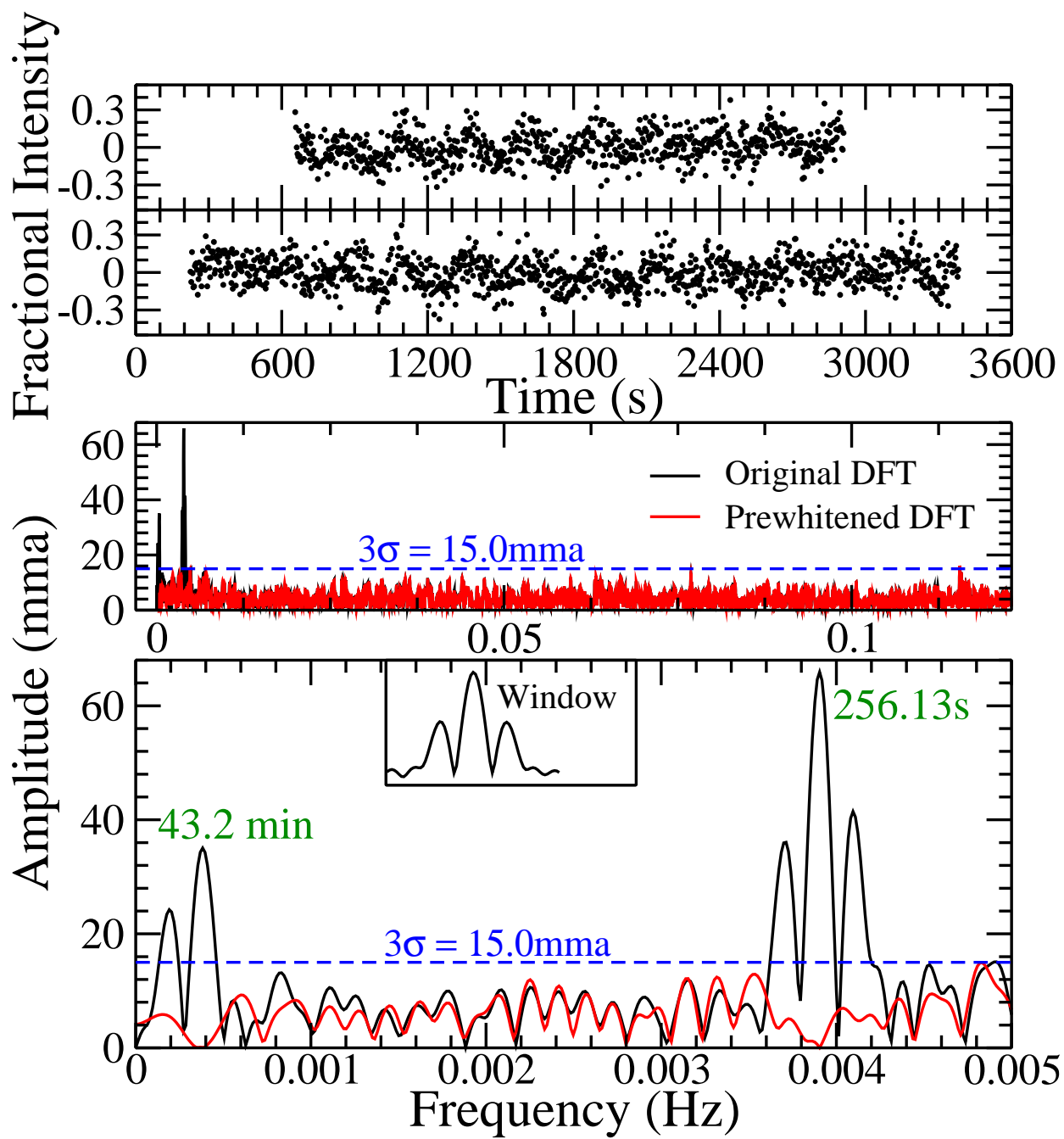}
\caption{COS intensity light curve 14 months after outburst with 3s bins (top), DFT (middle)
and an expansion of the low frequencies (bottom). Dashed line shows
the 3$\sigma$ noise limit found from the shuffling technique. Inset
on bottom shows the window function for the data. The 256 sec period is well-detected.}
\end{figure}

\clearpage
\begin{figure}
\figurenum {5}
\plotone{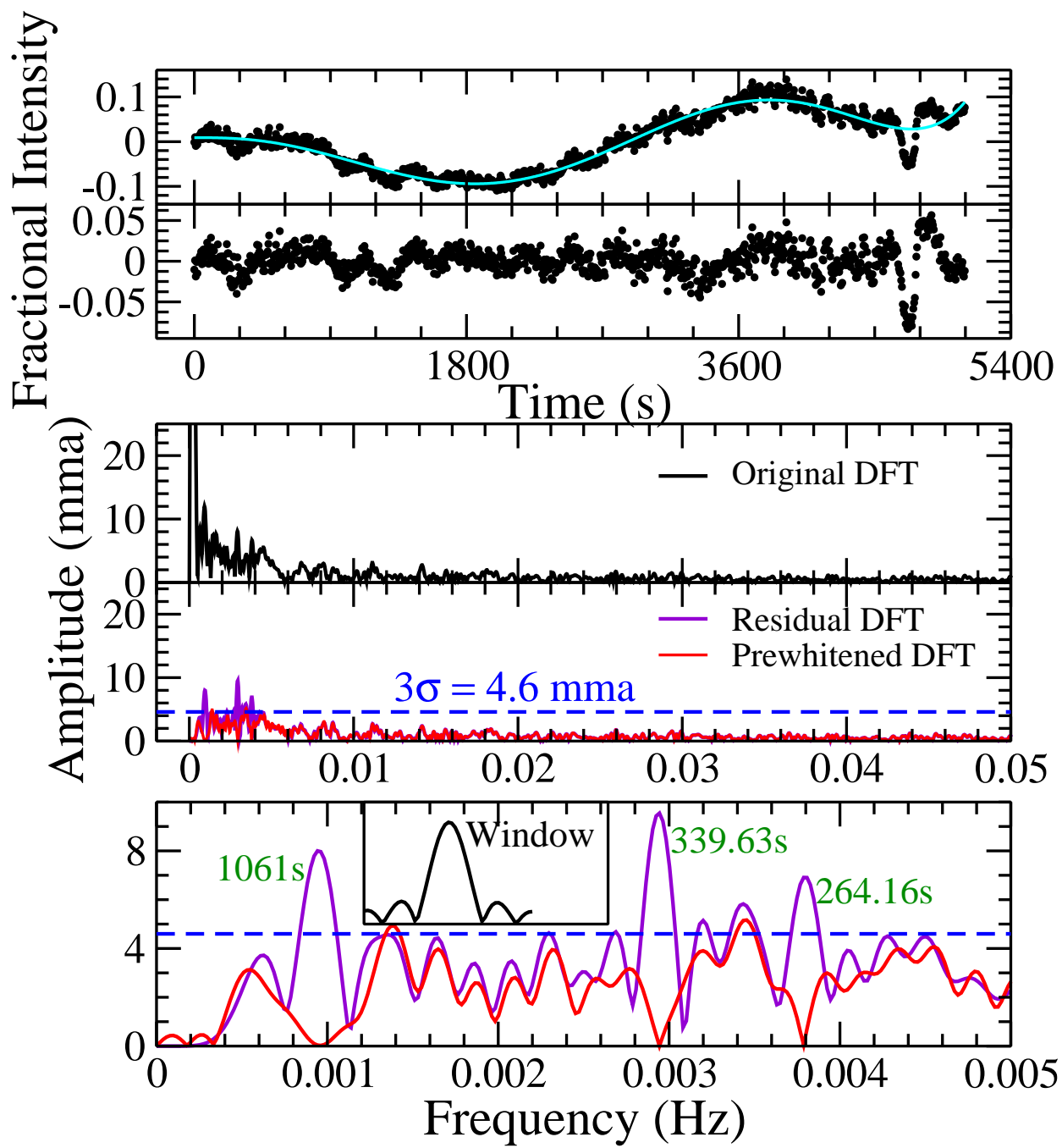}
\caption{APO 2010 November 8 data 1.5 months after outburst at magnitude 
17.7$\pm$0.2. Top panel shows original intensity light curve with 5s exposures; 
underneath is this light curve with a low order polynomial fit subtracted to 
remove the low frequency trend. Middle panels 3 and 4 show DFTs of the above 
light curves
while the bottom panel shows  
an expansion of the low frequencies in panel 4 and an inset with
 the window function. 
Dashed lines show
the 3$\sigma$ noise limit found from the shuffling technique. The pronounced dip near
the end of the light curve is the partial eclipse, with time listed in Table 2.
The 256 sec period is not evident but is likely masked by the other variability
due to the outburst.}
\end{figure}

\clearpage
\begin{figure}
\figurenum {6}
\plotone{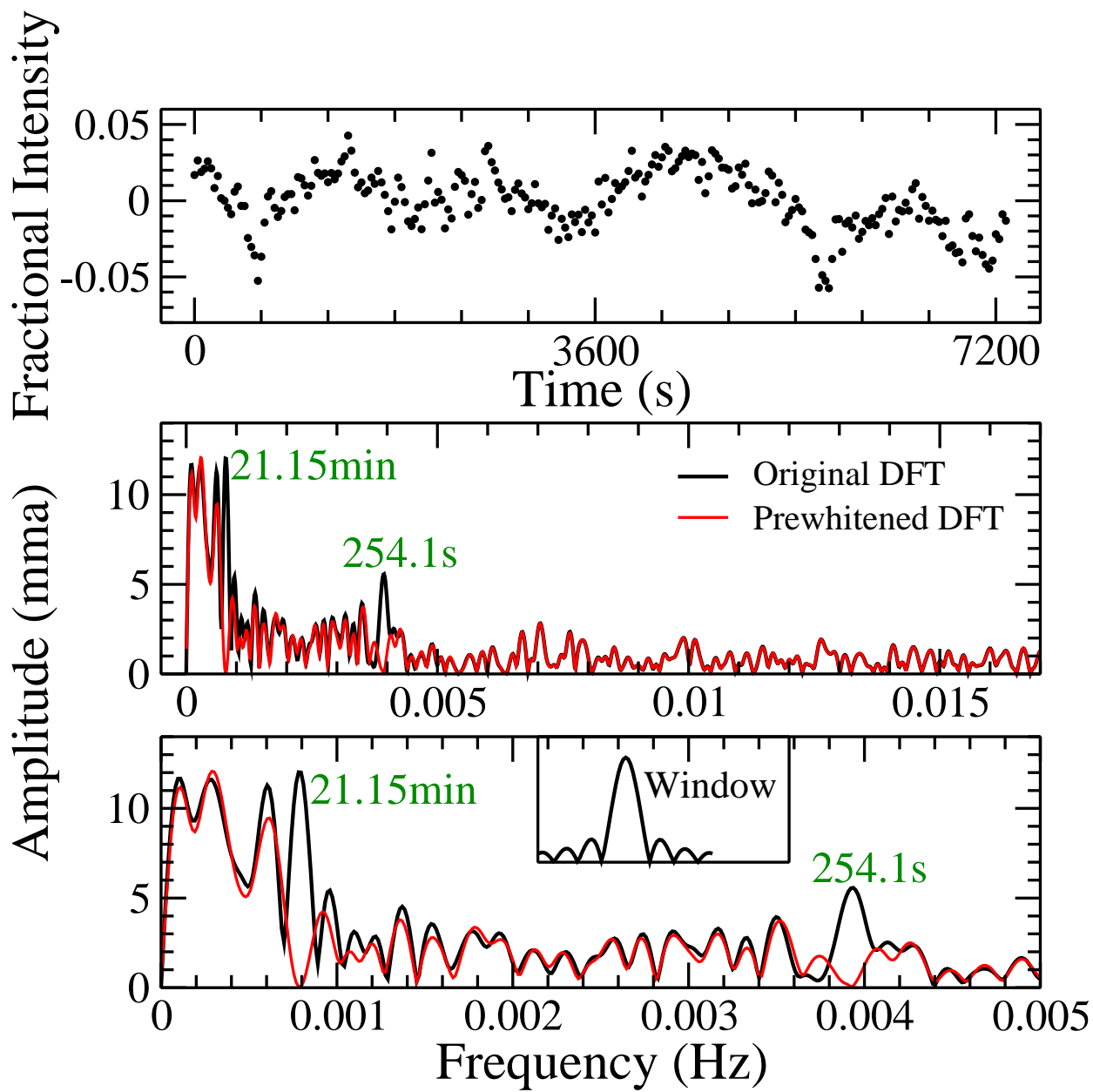}
\caption{Similar to Figure 4 showing APO data from 2011 November 5 
(14 months after outburst) at 
magnitude 17.8$\pm$0.1 with 30s exposures. Two partial eclipses are evident in
the light curve (Table 2). A signal consistent with the 256 sec period evident
in the COS data is detected. Due to the short length of the dataset, the 
shuffling technique could not produce a good value of 3$\sigma$ over the 
entire frequency range but an estimate for frequencies $>$ 0.005 Hz is 2.9 mma.}
\end{figure}
 
\clearpage
\begin{figure}
\figurenum {7}
\plotone{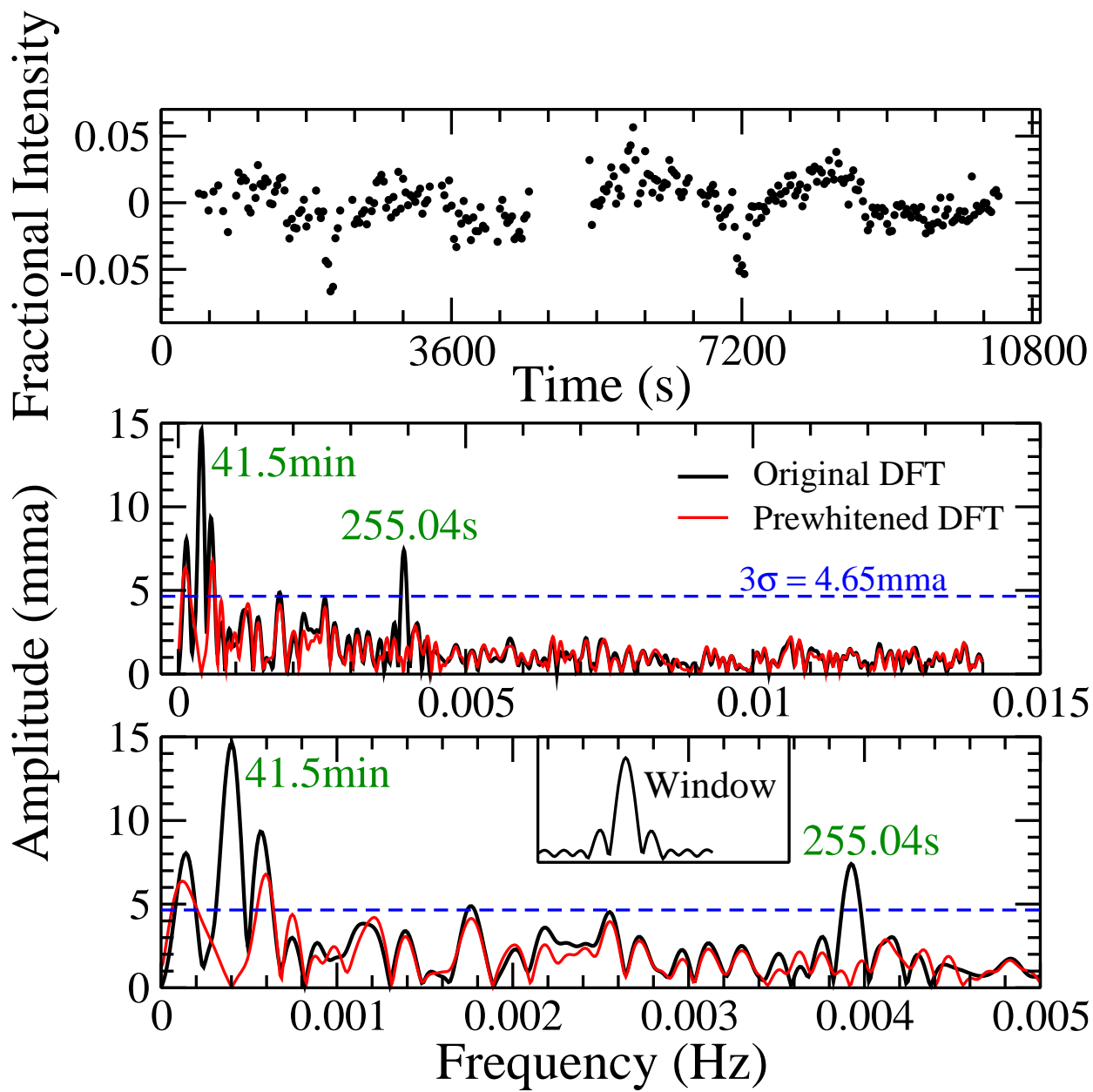}
\caption{Similar to Figure 4 with data from APO with 30-60s exposures on
2011 November 7. Mean magnitude is 17.5$\pm$0.1 and 2 partial eclipses are
present (Table 2). The 256 sec period is even more robustly detected in this
longer dataset.}
\end{figure}


\begin{references}
\reference{} Arras, P., Townsley, D. M. \& Bildsten, L. 2006, \apj, 643, L119
\reference{} Aviles, A. et al. 2010, \apj, 711, 389
\reference{} Brickhill, A. J. 1992, \mnras, 259, 519
\reference{} Bullock, E. et al. 2011,\aj, 141, 84
\reference{} Chote, P. \& Sullivan, D. J. 2013, ASPCS, in press
\reference{} G\"ansicke, B. T. \& Beuermann, K. 1996, \aap, 309, L47
\reference{} G\"ansicke, B. T., Szkody, P., Howell, S. B. \& Sion, E. M. 2005, \apj, 629, 451
\reference{} G\"ansicke, B. T. et al. 2009, \mnras, 397, 2170
\reference{} Gianninas, A., Bergeron, P. \& Fontaine, G. 2005, \apj, 631, 1100
\reference{} Godon, P. et al. 2006,\apj, 642, 1018
\reference{} Goldreich,P. \& Wu, Y. 1999, \apj, 511, 904
\reference{} Copperwheat, C. M. et al. 2009, \mnras, 393, 157
\reference{} Howell, S. B., Szkody, P. \& Cannizzo, J. 1995,\apj, 439, 337
\reference{} Hubeny, I. \& Lanz, T, 1995, \apj, 439, 875
\reference{} Kato, T. et al. 2009, \pasj, 61, 601
\reference{} Kato, T. et al. 2012, \pasj, 64, 21
\reference{} Montgomery, M. H. 2005, \apj, 633, 1142
\reference{} Mukadam, A. S. et al. 2011, \pasp, 123, 1423
\reference{} Nitta,A. et al. 2000, BaltA, 9, 97
\reference{} Patterson, J. et al. 1998, \pasp, 110, 403
\reference{} Pavlenko, E. et al. 2007, ASPCS, 372, 511
\reference{} Pavlenko, E. 2009, JPhCS, 172, 2071
\reference{} Pavlenko, E. et al. 2012, Mem. S. A. It. 172, 2071
\reference{} Piro, A. L., Arras, P. \& Bildsten, L. 2005, \apj, 628, 401
\reference{} Robinson, E. L. et al. 1995, \apj, 438, 908
\reference{} Silvestri, N. M. et al. 2012, \aj, 144, 84
\reference{} Sion, E. M. 1999, \pasp, 111,532
\reference{} Szkody, P., G\"ansicke, B. T., Sion, E. M., \& Howell, S. B. 2002, \apj, 574, 950
\reference{} Szkody, P. et al. 2003, \aj, 126, 1499
\reference{} Szkody, P. et al. 2006, \aj, 131, 973
\reference{} Szkody, P. et al. 2007, \apj, 658, 1188
\reference{} Szkody, P. et al. 2010, \apj, 710, 64
\reference{} Szkody, P. et al. 2012, \apj, 753, 158
\reference{} Uthas, H. et al. 2012, \mnras, 420, 379
\reference{} van Zyl, L. et al. 2004, \mnras, 350, 307
\reference{} Vican, L. et al. 2011, \pasp, 123, 1156
\reference{} Wu, Y. 2001, \mnras, 323, 248
\reference{} Zharikov, S. V. et al. 2008, \aap, 486, 505
\reference{} Zharikov, S. V. et al. 2013, \aap, 549, A77
\end{references}
\end{document}